\documentclass[aps,prd,showpacs,twocolumn]{revtex4}
\usepackage{amssymb}
\usepackage{amsmath}
\usepackage{color}
\usepackage[usenames,dvipsnames]{xcolor}
\usepackage{float}
\usepackage{accents}
\usepackage{graphicx}
\usepackage{epstopdf}
\usepackage{soul}
\usepackage[colorlinks=true,pdfstartview=FitV,linkcolor=blue,citecolor=blue,urlcolor=blue,breaklinks=true]{hyperref}

\begin{document}

\title{Gupta-Bleuler's {\ quantization of a} parity-odd CPT-even
electrodynamics of the standard model extension}
\author{R. Casana}
\email{rodolfo.casana@gmail.com}
\author{M. M. Ferreira Jr}
\email{manojr.ufma@gmail.com}
\affiliation{Universidade Federal do Maranh\~{a}o (UFMA), Departamento de F\'{\i}sica,
Campus Universit\'{a}rio do Bacanga, S\~{a}o Lu\'{\i}s - MA, 65080-805,
Brazil}
\author{F. E. P. dos Santos}
\email{fredegol@ibest.com.br}
\affiliation{Universidade Federal do Maranh\~{a}o (UFMA),}
\affiliation{Coordena\c c\~ao do Bacharelado Interdisciplinar em Ci\^encia e Tecnologia,
Campus Universit\'{a}rio do Bacanga, S\~{a}o Lu\'{\i}s - MA, 65080-805,
Brazil}

\begin{abstract}
{Following a successfully quantization scheme previously }{{%
developed {\ in Ref. \cite{GUPTAEVEN} for a}} parity-even {gauge
sector} {of the SME, we}} have established the Gupta-Bleuler {%
quantization {of a} {\ parity-odd} and CPT-even }electrodynamics of
{the }standard model extension (SME) {without recoursing to a
small photon mass regulator}. {Keeping the photons massless,} {%
{we have adopted the gauge fixing condition:}} $G(A_{\mu })=(\partial
_{0}+\kappa^{0j}\partial _{j}) (A_{0}+\kappa ^{0k}A_{k})+\partial _{i}A^{i}$%
. The{\ four} polarization vectors of the gauge field are {%
exactly} determined by solving an eigenvalue problem,{\ exhibiting
birefringent second order contributions in the Lorentz-violating parameters}%
. They allow to express the Hamiltonian in terms of annihilation and
creation operators whose positivity is guaranteed by imposing a weak
Gupta-Bleuler constraint, defining the physical states. Consequently, we
compute the field commutation relation which has been expressed in terms of
Pauli-Jordan functions modified by Lorentz violation whose light-cone
structures have allowed to analyze the microcausality issue.
\end{abstract}

\maketitle

\section{Introduction}

{Theories beyond the standard model of elementary particles have had
increasing interest in recent years}. {Among various }theoretical {%
approaches, the standard model extension (SME) \cite{Colladay} proposes the
possibility of {occurring spontaneous} breaking of Lorentz symmetry
in a fundamental theory at Planck {scale,} whose effects would be
mensurable at lower energy }{scenarios}. The SME is {constructed by
the {explicit} addition of Lorentz-violating (LV) terms {in}
all sectors of the standard {model,} {preserving} or not the
CPT symmetry}. In this context, the electromagnetic sector{\ has been
modified by CPT-odd \cite{Jackiw} and CPT-even \cite{Mewes} {LV terms.%
} Some aspects of its quantization and the related quantum electrodynamics}
have been studied in Refs. \cite{Soldati}. {\ The consistency analysis}
about causality, energy positivity and unitarity{\ were{\
accomplished for CPT-odd case Refs. \cite{Adam,Baeta} and for the
nonbirefringent CPT-even sector \cite{propag}}. The treatment of possible
infrared divergences issues, which would be present in a LV quantum
electrodynamics were conveniently solved in Ref. \cite{Massivephotons} by
giving a small mass for the photon field. An analysis {about }the
consistency of the nonbirefringent CPT-even and parity-odd quantum
electrodynamics was performed in Ref. \cite{Schreck}. The issue of the
quantization problem in birefringent Lorentz-violating quantum
electrodynamics was analyzed in \cite{Schreck2}.}

The problem of the Gupta-Bleuler (GB) {covariant }quantization {(in
the Lorenz gauge, $\partial ^{\mu }A_{\mu }=0$)} of the nonbirefringent
photonic sector of the SME was firstly{\ implemented at leading order
in the LV parameters} {in Ref. }\cite{Hohensee}. However, this useful
technique does not work beyond the leading order because the arising of the
birefringence phenomenon, whose effects cannot be eliminated by a simple
coordinate redefinition. In order to solve such a {problem}, {it was
considered a small mass for the photon field} {allowing\ the implementation}
{of the covariant quantization for this CPT-even LV electrodynamics}
\cite{colladay2014gupta}. {Nevertheless, the massless limit produces an
inconsistency because of the arising of an unavoidable incompleteness of the
polarization states \cite{colladay2014gupta,Colladaycov}}. {\ Such a
quantization scheme\ has also yielded a quantum Cherenkov radiation
evaluation} i{n a CPT-odd electrodynamics \cite{Colladaym} and, very
recently, its covariant quantization \cite{Colladaym2}.}

{Despite all these studies, the covariant quantization of a {Lorentz
violating birefringent electrodynamics, involving massless photons, }remains
{an open problem.}} Advances in this direction {were achieved}
{for} a {LV CPT-even electrodynamics} with anisotropic
parity-even coefficients \cite{GUPTAEVEN}. Such investigation also includes
small birefringence effects and {introduces} {a new} procedure
{to address the} {successful covariant quantization without
inclusion of photon mass}. The key factor {is the finding of a }%
compatible gauge fixing condition able to provide a consistent set of
polarization vectors.

The aim {of this work} is to apply the method first developed in Ref.
\cite{GUPTAEVEN} {to accomplish }{a consistent implementation of the
GB quantization for the {parity-odd sector of the CPT-even
electrodynamics of the SME,} {keeping the photon massless }during all
the procedure. {The goal was attained due to the choice} of the
following gauge} fixing condition, $G(A_{\mu })=(\partial _{0}+\kappa
^{0j}\partial _{j})(A_{0}+\kappa ^{0k}A_{k})+\partial _{i}A^{i}$, {a
LV} version of the usual Lorenz gauge. {We have evaluated }the
suitable polarization basis and determined the condition on the physical
states that assures {energy positivity}. Finally, we have analyzed
the microcausality and computed the Feynman propagator for this LV
electrodynamics.

\section{Aspects on the photon sector of the SME}

The CPT-even photonic sector of the SME includes a Lorentz-violating term
composed by the tensor $(k_{F}) ^{\mu \nu \alpha \beta },$ which possesses
the symmetries of the Riemann tensor and a null double trace, $(k_{F}) ^{\mu
\nu }{}_{\mu \nu }=0,$ implying 19 independent components. The corresponding
Lagrangian density,
\begin{equation}
\mathcal{L}=-\frac{1}{4}F_{\mu \nu }F^{\mu \nu }-\frac{1}{4}(k_{F}) ^{\mu
\nu \alpha \beta }F_{\mu \nu }F_{\alpha \beta },  \label{SMELagrangian}
\end{equation}%
is obviously invariant under U(1) gauge transformation. Here, $F_{\mu \nu
}=\partial _{\mu }A_{\nu }-\partial _{\nu }A_{\mu }$ is the Maxwell usual
tensor, $A_{\mu }$ is the Abelian gauge field. The second term is
responsible only by Lorentz-violation preserving CPT-symmetry (CPT-even).
From the 19 independents components of the background tensor, 10 are
birefringent and 9 are nonbirefringent. In another perspective, 11
components are parity-even and 8 are parity-odd. The parameterizations that
allow to distinguish and manipulate these components are described in Ref.
\cite{Mewes}.

In general, light birefringence in vacuum is a characteristic of the SME
gauge sector. Due to the observed magnitude of this effect for the light
coming from far distant galaxies, the LV terms yielding vacuum birefringence
undergo the most severe known constraints \cite{Mewes,data}. So, it is usual
to consider the {first order }birefringent LV coefficients as null.
The $9$ remaining nonbirefringent{\ }coefficients of $(k_{F})_{\mu \nu
\alpha \beta }$ can be parameterized by a symmetric and traceless tensor $%
\kappa ^{\mu \nu }$ \cite{kappamunu}, defined as
\begin{equation}
(k_{F})_{\mu \nu \alpha \beta }=\frac{1}{2}\left( \eta _{\mu \alpha }\kappa
_{\nu \beta }-\eta _{\mu \beta }\kappa _{\nu \alpha }+\eta _{\nu \beta
}\kappa _{\mu \alpha }-\eta _{\nu \alpha }\kappa _{\mu \beta }\right) ,
\end{equation}%
which allows to write the Lagrangian density (\ref{SMELagrangian}) as
\begin{equation}
\mathcal{L}=-\frac{1}{4}F_{\mu \nu }F^{\mu \nu }-\frac{1}{2}\kappa _{\nu
\rho }F^{\mu \nu }F_{\mu }{}^{\rho }.  \label{NBLagrangian}
\end{equation}%
{Here,} $\kappa _{00}$ and $\kappa _{ij}$ stand for isotropic and
anisotropic parity-even components, {while} the parity-odd sector are
represented by the $\kappa _{0i}$ coefficients.

\subsection{Gupta-Bleuler {quantization of a parity-odd CPT-even
electrodynamics}}

In a covariant quantization scheme, all (four) degrees of freedom of the
gauge field should be quantized in the same way \cite{Gupta,Bleuler}. The
Gupta-Bleuler procedure introduces in the Lagrangian density a specific term
breaking the local gauge invariance but not eliminating any gauge field
degree of freedom. However, GB quantization is not compatible with all
possible gauge conditions, for example, in Maxwell electrodynamics, the
Lorenz condition $(\partial _{\mu }A^{\mu })^{2}/2{\xi }$, in the Feynman
gauge ($\xi =1$) works fine \cite{NAKANISHI}. On the other hand, in a
Lorentz-violating electrodynamics, the Feynman gauge becomes incompatible
with the Gupta-Bleuler procedure \cite{colladay2014gupta,GUPTAEVEN}. Such
incompatibility, in the case of the CPT-even and parity-even LV
electrodynamics, was solved in Ref. \cite{GUPTAEVEN}. It was shown that the
Gupta-Bleuler procedure is compatible with the generalized gauge condition, $%
\partial _{\mu }A^{\mu }+\kappa ^{\mu \nu }\partial _{\mu }A_{\nu }=0$ (only
for $\kappa _{ij}\neq 0$), a kind of LV version of the Lorentz gauge.

{The goal} is to implement the Gupta-Bleuler quantization {of
the parity-odd and CPT-even electrodynamics}{\ of the SME {in
accordance with the technique developed in} Ref. \cite{GUPTAEVEN}.} Thus,
our analysis will be based in the following Lagrangian density:
\begin{equation}
\mathrm{{\mathcal{L}}}=-\frac{1}{4}F^{\mu \nu }F_{\mu \nu }+\kappa
_{0i}F^{ij}F^{0j}-\frac{1}{2\xi }\left( G\left[ A\right] \right) ^{2},
\label{lq1}
\end{equation}%
where $G\left[ A\right] $ is the suitable gauge fixing condition,
\begin{equation}
G\left[ A\right] =\left( \partial _{0}+\kappa ^{0j}\partial _{j}\right)
\left( A^{0}+\kappa _{0k}A^{k}\right) +\partial _{i}A^{i},  \label{gcc}
\end{equation}%
a modified relation involving the LV coefficients,\ compared to the usual
Lorenz gauge, i.e, $G\left[ A\right] =\partial _{\mu }A^{\mu }$. The
equation of motion for the gauge field, in the momentum space, reads
\begin{equation}
{O_{\mu \nu }}(p){A^{\nu }(}p)={0,}  \label{covMeq}
\end{equation}%
where the components of the tensor $O_{\mu \nu }$ are
\begin{align}
O_{00}& =\left( p_{0}-\vec{\kappa}\cdot \vec{p}\right) ^{2}-\left\vert \vec{p%
}\right\vert ^{2}, \\
O_{0i}& =\left[ \left( p_{0}-\vec{\kappa}\cdot \vec{p}\right)
^{2}-\left\vert \vec{p}\right\vert ^{2}\right] \kappa _{i}, \\
O_{i0}& =0, \\
O_{kj}& =-\left[ \left( p_{0}-\vec{\kappa}\cdot \vec{p}\right)
^{2}-\left\vert \vec{p}\right\vert ^{2}-\left( \vec{\kappa}\cdot \vec{p}%
\right) ^{2}\right] \delta _{jk}, \\
& -\left( \vec{\kappa}\cdot \vec{p}\right) \left[ \kappa _{k}p_{j}+\kappa
_{j}p_{k}\right] +\left\vert \vec{p}\right\vert ^{2}\kappa _{k}\kappa _{j}.
\notag
\end{align}%
where we have defined $\vec{\kappa}\cdot \vec{p}=\kappa _{i}p_{i}$ wich
vector $\kappa _{i}=\left( \kappa _{01},\kappa _{02},\kappa _{03}\right) $.

To obtain non trivial solutions of Eq. (\ref{covMeq}), the determinant of
the tensor $O{_{\mu \nu }}(p)$ must be zero,
\begin{equation}
\det O{_{\mu \nu }}(p)=\boxminus ^{2}(p)\mathrm{\boxplus }(p)\mathrm{%
\boxtimes }(p)=0.
\end{equation}%
Under the gauge fixing condition (\ref{gcc}), we obtain the following
dispersion relations:%
\begin{align}
\boxminus (p)& =\left( p_{0}-\vec{\kappa}\cdot \vec{p}\right)
^{2}-\left\vert \vec{p}\right\vert ^{2},  \label{dp1} \\
\mathrm{\boxplus }(p)& =\left( p_{0}-\vec{\kappa}\cdot \vec{p}\right)
^{2}-\left( 1+\left\vert \vec{\kappa}\right\vert ^{2}\right) \left\vert \vec{%
p}\right\vert ^{2},  \label{dp2} \\
\mathrm{\boxtimes }(p)& =\left( p_{0}-\vec{\kappa}\cdot \vec{p}\right)
^{2}-\left\vert \vec{p}\right\vert ^{2}-\left( \vec{\kappa}\cdot \vec{p}%
\right) ^{2}.  \label{dp3}
\end{align}%
The first one, $\boxminus (p)=0$, having multiplicity 2, is a nonphysical
dispersion relation. The two others, $\boxplus (p)=0$ and $\boxtimes (p)=0,$
correspond to the physical dispersion relations. Given a gauge condition,
the Gupta-Bleuler quantization can be implemented \cite{GUPTAEVEN} if the
dimension of the null space of the matrix operator $O^{\mu \nu }$ is equal
to its respective multiplicity, when the particular dispersion relation\ is
fulfilled.

We verify that the physical dispersion relations have $\dim \mathrm{Null}%
\left( O^{\mu \nu }\big\vert_{\mathrm{\boxplus }(p)=0}\right) =1$ and $\dim
\mathrm{Null}\left( O^{\mu \nu }\big\vert_{\mathrm{\boxtimes }(p)=0}\right)
=1$, revealing the existence of one polarization vector for each dispersion
relation. The nonphysical dispersion relation $\boxminus (p) $ provides $%
\dim \mathrm{Null}\left( O^{\mu \nu }\big\vert_{\boxminus (p)=0}\right) =2$,
being associated with two polarization vectors. Hence, it is guaranteed that
we can find a total of four polarization vectors. Thus, the gauge condition (%
\ref{gcc}) fulfills all conditions proposed in Ref. \cite{GUPTAEVEN} for
implementing the Gupta-Bleuler quantization.

In order to write the canonical Hamiltonian density, after an integration by
parts in (\ref{lq1}), we find the canonical conjugate momentum components,
\begin{align}
\pi _{0}& =-\left( \partial _{0}-\kappa _{i}\partial _{i}\right) \left(
A_{0}-\kappa _{j}A_{j}\right) , \\
\pi _{j}& =-\left( \partial _{0}-\kappa _{i}\partial _{i}\right) A_{j} \\
& -\kappa _{j}\left( \partial _{0}-\kappa _{i}\partial _{i}\right) \left(
A_{0}-\kappa _{j}A_{j}\right) .  \notag
\end{align}

{These canonical momentum} allow to find the following Hamiltonian
density:%
\begin{align}
{\mathrm{{\mathcal{H}}}}& ={-\frac{1}{2}\pi _{0}\pi _{0}-\pi _{0}\kappa
_{i}\pi _{i}+\frac{1}{2}\pi _{0}\kappa _{i}\kappa _{i}\pi _{0}}  \notag \\
& {+\frac{1}{2}\pi _{j}\pi _{j}-\pi _{j}\kappa _{i}\partial _{i}A_{j}+\kappa
_{i}\pi _{0}\partial _{i}A_{0}}  \notag \\
& -{\frac{1}{2}\partial _{i}\left( A{_{0}-}\kappa _{j}{A_{j}}\right)
\partial _{i}\left( A{_{0}-}\kappa _{k}{A_{k}}\right) }  \notag \\
& -\kappa _{i}{\kappa _{j}\partial _{i}A_{k}}\partial _{k}A_{j}+\frac{1}{2}%
\left[ {\partial _{i}A_{j}}\partial _{i}A_{j}\right.  \notag \\
& ~\left. +{\kappa _{i}}\kappa _{j}({\partial _{j}A_{k}}\partial
_{i}A_{k}+\kappa _{i}{\kappa _{j}\partial _{k}A_{i}}\partial _{k}A_{j})%
\right] .  \label{Hcov}
\end{align}%
The canonical commutation relations read%
\begin{equation}
\left[ A^{\mu }(t,\vec{x}),\pi _{\nu }(t,\vec{y})\right] =i\delta _{\nu
}^{\mu }\delta ^{3}(\vec{x}-\vec{y}),  \label{canonical}
\end{equation}%
\begin{equation}
\left[ A^{\mu }(t,\vec{x}),A_{\nu }(t,\vec{y})\right] =\left[ \pi ^{\mu }(t,%
\vec{x}),\pi _{\nu }(t,\vec{y})\right] =0.  \label{canonicalnull}
\end{equation}

{We now can observe that }$\left[ A^{0},G(A_{\mu })\right] \neq 0$%
{\ is a necessary condition for the GB quantization. However, it
fails at operator level when }$G(A_{\mu })=0${, as it occurs for some
gauge choices, like the Lorenz gauge \cite{colladay2014gupta}. To avoid this
problem, one chooses a gauge condition, namely Eq. {(\ref{gcc}), which}
provides a set of four linearly independent polarization vectors.}

In the next step, we can propose a solution for equation (\ref{covMeq}) in
terms of wave plane expansion,%
\begin{equation}
A^{\mu }=\int \sum_{\lambda =0}^{3}\widehat{d^{3}\vec{p}}_{(\lambda )}\left(
a_{(\lambda )}e^{-ip_{(\lambda )}\cdot x}+a_{(\lambda )}^{\dag
}e^{ip_{(\lambda )}\cdot x}\right) \varepsilon _{(\lambda )}^{\mu },
\label{covPW}
\end{equation}%
where $\lambda $ denotes a polarization mode, $\widehat{d^{3}\vec{p}}%
_{(\lambda )}=\left[ \left( 2\pi \right) ^{3}2C_{(\lambda )}\right]
^{-1/2}d^{3}\vec{p}$, $C_{(\lambda )}$ is a normalization factor, $%
a_{(\lambda )}$ and $a_{(\lambda )}^{\dag }$ are the respective annihilation
and creation operators, and $\varepsilon _{(\lambda )}^{\mu }$ is the
polarization vector.

To satisfy the canonical commutation relations (\ref{canonical}), (\ref%
{canonicalnull}), the creation and annihilation operators must satisfy the
following relations:%
\begin{equation}
\left[ a_{(\lambda)}^{\dag}(p) ,a_{\left( \lambda^{\prime }\right) }\left(
q\right) \right] =g_{\lambda\lambda^{\prime}}\delta ^{3}\left( p-q\right) .
\label{nonullcomutationp}
\end{equation}%
\begin{equation}
\left[ a_{(\lambda)}(p) ,a_{\left( \lambda^{\prime}\right) }\left( q\right) %
\right] =\left[ a_{(\lambda)}^{\dag}(p) ,a_{\left( \lambda^{\prime}\right)
}^{\dag}\left( q\right) \right] =0.
\end{equation}

The polarization vectors are computed from the eigenvalue equation,
\begin{equation}
O_{\alpha \beta }\varepsilon _{(\lambda )}^{\beta }=\alpha _{(\lambda
)}\left( g_{\alpha \beta }+g_{\alpha 0}\kappa _{0\beta }\right) \varepsilon
_{(\lambda )}^{\beta },
\end{equation}%
which provides
\begin{align}
\varepsilon _{(0)}^{\mu }\left( \vec{p}\right) & =\left[ 1,0,0,0\right]
\label{pp1} \\
\varepsilon _{(1)}^{\mu }\left( \vec{p}\right) & =\left[ \vec{\kappa}\cdot
\vec{\varepsilon}_{(1)}{\left( \vec{p}\right) },\vec{\varepsilon}_{(1)}{%
\left( \vec{p}\right) }\right]  \label{pp2} \\
\varepsilon _{(2)}^{\mu }\left( \vec{p}\right) & =\left[ \vec{\kappa}\cdot
\vec{\varepsilon}_{(2)}{\left( \vec{p}\right) },\vec{\varepsilon}_{(2)}{%
\left( \vec{p}\right) }\right]  \label{pp3} \\
\varepsilon _{\left( 3\right) }^{\mu }\left( \vec{p}\right) & =\left[ \vec{%
\kappa}\cdot \vec{\varepsilon}_{(3)}{\left( \vec{p}\right) },\vec{\varepsilon%
}_{(3)}{\left( \vec{p}\right) }\right]  \label{pp4}
\end{align}%
with the eigenvalues $\alpha _{(\lambda )}$ given by
\begin{equation}
\alpha _{(0)}=\alpha _{(3)}=\boxminus (p),\;\alpha _{(1)}=\mathrm{\boxtimes }%
(p),\;\alpha _{(2)}=\mathrm{\boxplus }(p).  \label{alpha}
\end{equation}%
The three vectors $\vec{\varepsilon}_{(1)}${, }$\vec{\varepsilon}_{(2)}$ and
$\vec{\varepsilon}_{(3)}$ are given by
\begin{align}
\vec{\varepsilon}_{(1)}& =\frac{\vec{p}\times \vec{\kappa}}{\left\vert \vec{p%
}\times \vec{\kappa}\right\vert },  \label{pol1} \\
\vec{\varepsilon}_{(2)}& =\frac{\left\vert \vec{p}\right\vert }{\left\vert
\vec{p}\times \vec{\kappa}\right\vert }\left( \vec{\kappa}-\frac{(\vec{\kappa%
}\cdot \vec{p})}{\left\vert \vec{p}\right\vert ^{2}}\vec{p}\right) ,
\label{pol2} \\
\vec{\varepsilon}_{(3)}& =\frac{\vec{p}}{\left\vert \vec{p}\right\vert }.
\label{pol3}
\end{align}%
\qquad {It is worthwhile to note that the set of polarization vectors
{should be calculated separately when} the vectors $\vec{p}$ and $%
\vec{\kappa}$ are parallel. In spite of that, our formalism works {%
properly} for this configuration {as well }(see Sec.\ref{app}).}

The polarization vectors verify the normalization condition%
\begin{equation}
g_{\mu \nu }\left( \delta _{\alpha }^{\mu }+\delta _{0}^{\mu }\kappa
_{0\alpha }\right) \left( \delta _{\beta }^{\nu }+\delta _{0}^{\nu }\kappa
_{0\beta }\right) \varepsilon _{(\lambda )}^{\alpha }\left( \vec{p}\right)
\varepsilon _{\left( \lambda ^{\prime }\right) }^{\beta }\left( \vec{p}%
\right) =g_{\lambda \lambda ^{\prime }},
\end{equation}%
and the completeness condition
\begin{equation}
\sum_{\lambda =0}^{3}g_{\lambda \lambda }\varepsilon _{(\lambda )}^{\alpha
}\varepsilon _{(\lambda )}^{\beta }=\left( \delta _{\mu }^{\alpha }-\delta
_{0}^{\alpha }\kappa _{0\mu }\right) g^{\mu \nu }\left( \delta _{\nu
}^{\beta }-\delta _{0}^{\beta }\kappa _{0\nu }\right) .  \label{completness}
\end{equation}%
Consequently, the energy $E_{(\lambda )}$ of every polarization mode is
obtained from the respective eigenvalue $\alpha _{(\lambda )}=0$,%
\begin{align}
{E_{(0)}}& {=}{E_{\left( 3\right) }=\vec{\kappa}\cdot \vec{p}}+{\left\vert
\vec{p}\right\vert ,} \\
E{_{(1)}}& ={\vec{\kappa}\cdot \vec{p}+\left\vert \vec{p}\right\vert \sqrt{%
1+\left\vert \vec{\kappa}\right\vert ^{2}},} \\
{E_{(2)}}& {=}{\vec{\kappa}\cdot \vec{p}+\left\vert \vec{p}\right\vert \sqrt{%
1+\frac{\left( \vec{\kappa}\cdot \vec{p}\right) ^{2}}{\left\vert \vec{p}%
\right\vert ^{2}}}.}
\end{align}%
The two last energies are in exactly concordance with the physical
dispersion relations obtained in Ref. \cite{propag}, which shows the
consistency of this procedure of the dynamical respects of the theory.

The normalization factors $C_{(\lambda )}$ are conveniently defined as
\begin{equation}
C_{(\lambda )}=E_{(\lambda )}-\vec{\kappa}\cdot \vec{p}.
\end{equation}

The goal of our choice for the polarization basis (\ref{pp1})-(\ref{pp4}) is
to express the quantum Hamiltonian as an explicit sum of the contributions
of each polarization mode, as required,\
\begin{equation}
H=-\sum_{\lambda =0}^{3}\int d^{3}\vec{p}~E_{(\lambda )}N_{(\lambda
)}g_{\lambda \lambda ^{\prime }},  \label{covQH}
\end{equation}%
where\ $N_{(\lambda )}=a_{(\lambda )}^{\dag }a_{(\lambda )}$ is the number
operator for the polarization mode $\lambda $. Despite the fact that the
Hamiltonian can be expressed in a simple form, it is not positive definite.

Such as it happens in usual quantum electrodynamics, at operator level, the
condition $G\left[ A\right] =0$ is not compatible with commutation relations
(\ref{canonical}) and (\ref{canonicalnull}). Another problem is that the
operators $a_{(0)}^{\dag }$ and $a_{(0)}$ \ satisfy\ a commutation relation
with wrong signal which leads us to negative norm\ states. All these
problems are solved\ by imposing that the physical states $\left\vert
\varphi _{phys}\right\rangle $\ must satisfy the condition,%
\begin{equation}
{\left\langle \varphi _{phys}\right\vert \left( \partial _{0}-\kappa
_{j}\partial _{j}\right) \left( A_{0}-\kappa _{k}A_{k}\right) -\partial
_{i}A_{i}\left\vert \varphi _{phys}\right\rangle }=0,  \label{GBstrong}
\end{equation}%
in total analogy with the Gupta-Bleuler formalism. The last operatorial
condition is very strong and it is sufficient to impose a weaker operator
condition
\begin{equation}
\left[ \left( \partial _{0}-\kappa _{j}\partial _{j}\right) {\left(
A_{0}^{\left( +\right) }-\kappa _{k}A_{k}^{\left( +\right) }\right) }-{%
\partial _{i}A_{i}^{\left( +\right) }}\right] {\left\vert \varphi
_{phys}\right\rangle }=0,  \label{GBweak}
\end{equation}%
to select the physical states. Here, we remember the gauge field was
decomposed in positive and negative frequencies, $A_{\mu }=A_{\mu }^{\left(
+\right) }+A_{\mu }^{\left( -\right) }$, respectively. It is easy show that,
if (\ref{GBweak}) is true, then (\ref{GBstrong}) will also be true. By
implementing the weaker condition (\ref{GBweak}) in the plane-wave expansion
(\ref{covPW}) of the gauge field, we obtain%
\begin{align}
0& =\int \sum_{\lambda =0}^{3}\widehat{d^{3}\vec{p}}_{(\lambda
)}~e^{-ip_{(\lambda )}\cdot x}  \label{GBweak2} \\
& \hspace{-0.7cm}\times \left[ \left( p_{0(\lambda )}-\kappa
_{j}p_{j}\right) \left( \varepsilon _{0(\lambda )}-\kappa _{k}\varepsilon
_{k(\lambda )}\right) -p_{i}\varepsilon _{i(\lambda )}\right] a_{(\lambda )}{%
\left\vert \varphi _{phys}\right\rangle }.  \notag
\end{align}%
By using our polarization basis (\ref{pp1})-(\ref{pp4}), is easy to show that%
\begin{align}
\left( p_{0(0)}-\kappa _{j}p_{j}\right) \left( \varepsilon _{0(0)}-\kappa
_{k}\varepsilon _{k(0)}\right) -p_{i}\varepsilon _{i(0)}& =\left\vert \vec{p}%
\right\vert , \\
\left( p_{0(1)}-\kappa _{j}p_{j}\right) \left( \varepsilon _{0(1)}-\kappa
_{k}\varepsilon _{k(1)}\right) -p_{i}\varepsilon _{i(1)}& =0, \\
\left( p_{0(2)}-\kappa _{j}p_{j}\right) \left( \varepsilon _{0(2)}-\kappa
_{k}\varepsilon _{k(2)}\right) -p_{i}\varepsilon _{i(2)}& =0, \\
\left( p_{0(3)}-\kappa _{j}p_{j}\right) \left( \varepsilon _{0(3)}-\kappa
_{k}\varepsilon _{k(3)}\right) -p_{i}\varepsilon _{i(3)}& =-\left\vert \vec{p%
}\right\vert .
\end{align}%
By implementing in (\ref{GBweak2}), we obtain,%
\begin{equation}
\int \frac{d^{3}\vec{p}}{\sqrt{\left( 2\pi \right) ^{3}2C_{(0)}}}i\left\vert
\vec{p}\right\vert e^{ip_{(0)}\cdot x}\left[ a_{(0)}-a_{\left( 3\right) }%
\right] {\left\vert \varphi _{phys}\right\rangle }=0,
\end{equation}%
which yields the following constraint on the physical states:
\begin{equation}
\left[ a_{(0)}-a_{\left( 3\right) }\right] {\left\vert \varphi
_{phys}\right\rangle }=0.  \label{cxcx}
\end{equation}%
It allows to show that the expectation value of number operator of the
scalar\ and longitudinal modes are equal,
\begin{equation}
{\left\langle \varphi _{f}\right\vert }N_{(0)}{\left\vert \varphi
_{f}\right\rangle }={\left\langle \varphi _{f}\right\vert }N_{\left(
3\right) }{\left\vert \varphi _{f}\right\rangle .}  \label{GBcondictioN}
\end{equation}%
Then, the condition (\ref{cxcx}) solves the problem concerning the negative
norm states and the hamiltonian (\ref{covQH}) becomes positive definite for
physical states.

\subsection{The Pauli-Jordan function, microcausality and Feynman propagator}

Once we have successfully quantized this Lorentz-violating electrodynamics,
we can also compute the covariant commutation relation for the gauge field.
By using the plane wave expansion (\ref{covPW}) for the gauge field, we
obtain
\begin{align}
\left[ A^{\mu }(x),A^{\nu }(y)\right] & =-T_{(1)}^{\mu \nu }(i\partial
)i\Delta _{(1)}(x-y)  \label{cmc} \\[0.06in]
& -T_{(2)}^{\mu \nu }(i\partial )i\Delta _{(2)}(x-y)  \notag \\[0.15cm]
& -\left[ T_{(0)}^{\mu \nu }(i\partial )-T_{\left( 3\right) }^{\mu \nu
}(i\partial )\right] i\Delta _{(0)}(x-y),  \notag
\end{align}%
where we have introduced the projectors $T_{(\lambda )}^{\mu \nu }(i\partial
)$, which in the momentum space are defined as
\begin{equation}
T_{(\lambda )}^{\mu \nu }(p) =\varepsilon _{(\lambda )}^{\mu }(p)
\varepsilon _{(\lambda )}^{\nu }(p) .
\end{equation}%
On the other hand, the functions $\Delta _{(\lambda )}(x)$ represent the
Pauli-Jordan functions modified by Lorentz violation.

The nonnull components of the projectors $T_{(\lambda )}^{\mu \nu }(p) $ are
\begin{align}
T_{(0)}^{00}(p) & =1,~T_{ij(1)}(p) =\frac{\left( \vec{p}\times \vec{\kappa}%
\right) _{i}\left( \vec{p}\times \vec{\kappa}\right) _{j}}{\left\vert \vec{p}%
\times \vec{\kappa}\right\vert ^{2}}, \\[0.15cm]
T_{00(2)}(p) & =\frac{\left\vert \vec{\kappa}\times \vec{p}\right\vert ^{2}}{%
\left\vert \vec{p}\right\vert ^{2}},~T_{i0(2)}(p) =\kappa _{i}-\frac{\left(
\vec{\kappa}\cdot \vec{p}\right) }{\left\vert \vec{p}\right\vert ^{2}}p_{i},
\\[0.15cm]
T_{ij(2)}(p) & =\frac{\left\vert \vec{p}\right\vert ^{2}}{\left\vert \vec{%
\kappa}\times \vec{p}\right\vert ^{2}}\left( \kappa _{i}\kappa _{j}+\frac{%
\left( \vec{\kappa}\cdot \vec{p}\right) ^{2}}{\left\vert \vec{p}\right\vert
^{4}}p_{i}p_{j}\right. \\
& \hspace{1.75cm}\left. -\frac{\left( \vec{\kappa}\cdot \vec{p}\right) }{%
\left\vert \vec{p}\right\vert ^{2}}\left( p_{i}\kappa _{j}+p_{j}\kappa
_{i}\right) \right) ,  \notag \\[0.15cm]
T_{00(3)}(p) & =\frac{\left( \vec{\kappa}\cdot \vec{p}\right) ^{2}}{%
\left\vert \vec{p}\right\vert ^{2}},~T_{0i(3)}(p) ={-}\frac{\vec{\kappa}%
\cdot \vec{p}}{\left\vert \vec{p}\right\vert ^{2}}p_{i}, \\[0.15cm]
T_{ij(3)}(p) & =\frac{p_{i}p_{j}}{\left\vert \vec{p}\right\vert ^{2}}.
\end{align}%
The LV Pauli-Jordan functions $\Delta _{(\lambda )}(x)$ are given by
\begin{align}
\Delta _{(0)}(x)& =-\frac{\varepsilon (x_{0})}{2\pi }\delta \left(
x^{2}+2x_{0}\left( \vec{\kappa}\cdot \vec{x}\right) +\left( \vec{\kappa}%
\cdot \vec{x}\right) ^{2}\right) ,  \label{generalPJ0} \\[0.15cm]
\Delta _{(1)}(x)& =-\frac{\varepsilon (x_{0})}{2\pi }\frac{\delta \left(
x^{2}+2x_{0}\left( \vec{\kappa}\cdot \vec{x}\right) \right) }{\left(
1+\left\vert \vec{\kappa}\right\vert ^{2}\right) ^{1/2}},  \label{generalPJ1}
\\[0.15cm]
\Delta _{(2)}(x)& =-\frac{\varepsilon (x_{0})}{2\pi }\frac{\delta \left(
x^{2}+2x_{0}\left( \vec{\kappa}\cdot \vec{x}\right) -\left\vert \vec{\kappa}%
\times \vec{x}\right\vert ^{2}\right) }{\left( 1+\left\vert \vec{\kappa}%
\right\vert ^{2}\right) ^{-1/2}}.  \label{generalPJ2}
\end{align}%
These functions, in absence of Lorentz violation, become the usual
Pauli-Jordan function. On the other hand, by considering only the first
order LV contribution, they become exactly equal. This is expected because
the model (\ref{NBLagrangian}) is nonbirefringent at first order in Lorentz
violation.

The causal structure of the commutator is full determined by the functions $%
\Delta _{(\lambda )}(x),$ so there are three deformed light cones, one for
each dispersion relation. These light-cones degenerate to only one when we
consider only the\ first-order contributions.
\begin{figure}[]
\centering\includegraphics[width=8cm]{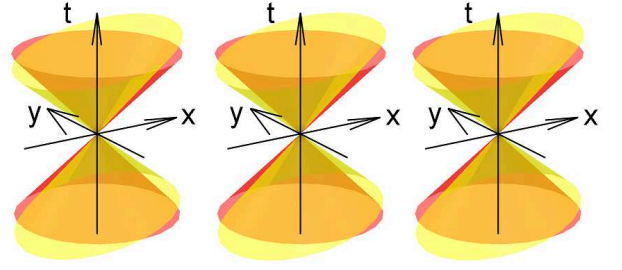}
\caption{Light-cone structure in a coordinate system where $\vec{k}=(1,0,0)$
and $\vec{x}=(x,y,0)$ for the functions (\protect\ref{generalPJ0}), (\protect
\ref{generalPJ1}) and (\protect\ref{generalPJ2}), respectively. For this
case, we can see the extreme deformation of light-cone in comparison to
usual one.}
\label{coneodd}
\end{figure}

The Fig. \ref{coneodd} represents the light-cone structures in a coordinate
system where we have fixed the LV vector $\vec{\kappa}$ along the $x$-axis, $%
\vec{\kappa}=(1,0,0)$ and $\vec{x}=(x,y,0)$. We can see the extreme
deformation of light cone in comparison to usual one. On the other hand, for
the cases $\vec{x}=(x,0,z)$ and $\vec{x}=(0,y,z)$ the light-cone is only
slightly different to the usual.

For microcausality analysis, we need that the functions $\Delta _{(\lambda
)}(x)$ vanish for space-like vectors $x^{2}<0$. To show it, we will use the
Lorentz-invariant frame where $x^{\mu }=\left( 0,\vec{x}\right) $. First {%
considering} the LV Paul-Jordan function (\ref{generalPJ0}), we note the
argument verifies the causality condition,
\begin{equation}
-\left\vert \vec{x}\right\vert ^{2}\left( 1-\left\vert \vec{\kappa}%
\right\vert ^{2}\cos ^{2}\theta \right) \,<0,
\end{equation}%
whenever $\left\vert \vec{\kappa}\right\vert <1$. The argument of the
modified Pauli-Jordan function (\ref{generalPJ1}) also verifies the
causality condition in the Lorentz-invariant frame,
\begin{equation}
-\left\vert \vec{x}\right\vert ^{2}<0.
\end{equation}%
The argument of the third LV Pauli-Jordan function (\ref{generalPJ2})
becomes
\begin{equation}
-\left\vert \vec{x}\right\vert ^{2}\left( 1+\left\vert \vec{\kappa}%
\right\vert ^{2}\sin ^{2}\theta \right) <0,
\end{equation}%
for all values of $\left\vert \vec{\kappa}\right\vert $. Hence, we verify
the functions $\Delta _{(\lambda )}(x)$ preserve the light-cone structure
whenever $\left\vert \vec{\kappa}\right\vert $ is sufficiently small.
Consequently, the functions $\Delta _{\left( \lambda \right) }(x)$ vanishes
for spacelike vectors, assuring the microcausality. We thus conclude that
the microcausality is preserved in this CPT-even and parity-odd LV
electrodynamics whenever $\left\vert \vec{\kappa}\right\vert $ is
sufficiently small. This microcausality analysis is in total accordance with
the one performed in Ref. \cite{Schreck} for this same model.

The Feynman propagator can be also computed%
\begin{equation}
{\left\langle 0\right\vert }TA^{\mu }(x)A^{\nu }\left( y\right) {\left\vert
0\right\rangle }=i\int \frac{d^{4}p}{\left( 2\pi \right) ^{4}}~\tilde{D}%
^{\mu \nu }(p) e^{-ip\cdot (x-y)},  \label{VVA}
\end{equation}%
where $\tilde{D}^{\mu \nu }(p) $ is given by%
\begin{equation}
\tilde{D}^{\mu \beta }(p) =-\sum_{\lambda =0}^{3}\frac{g_{\lambda \lambda
}\varepsilon _{\left( \lambda \right) }^{\mu }\left( \vec{p}\right)
\varepsilon _{\left( \lambda \right) }^{\nu }\left( \vec{p}\right) \left(
\delta _{\nu }^{\beta }+\delta _{0}^{\beta }\kappa _{0\nu }\right) }{\alpha
_{\left( \lambda \right) }+i\varepsilon },  \label{VVAp}
\end{equation}%
with $\alpha _{(\lambda )}$ defined in Eq. (\ref{alpha}). The tensor $\tilde{%
D}^{\mu \nu }(p) $ satisfies the following equation,%
\begin{equation}
O^{\mu}{}_{\alpha}(p) \tilde{D}^{\alpha \nu }(p) =-g^{\mu\nu }.
\label{1.108)}
\end{equation}%
Despite the propagator (\ref{VVAp}) presents a slightly different tensor
structure, the physical poles are the same found in Ref. \cite{propag}.

\section{Failure of the gauge condition $G(A_{\protect\mu })=\partial _{%
\protect\mu }A^{\protect\mu }+\protect\kappa _{\protect\mu \protect\nu %
}\partial ^{\protect\mu }A^{\protect\nu }$ in the CPT-even and parity-odd
case\label{app}}

{In this section, we show the incompatibility of the gauge fixing
condition, } $G(A_{\mu })=\partial _{\mu }A^{\mu }+\kappa _{\mu \nu
}\partial ^{\mu }A^{\nu },${\ with the implementation of the
Gupta-Bleuler quantization for the parity-odd and CPT-even electrodynamics.
For it, we consider, without loss of\ generality, the case when the 3D
vectors }$p_{i}${\ and }$\kappa _{i}${\ are parallel and along
the positive }$z${-axis. In this gauge, the expression for the
operator }$O^{\mu \nu }${\ defining the equation of motion of the
gauge field, }$O^{\mu \nu }(p)A_{\nu }(p)=0${, in the momentum space,
reads
\begin{widetext}
\begin{equation}
O^{\mu\nu}(p)=\left(
\begin{array}{cccc}
{\left\vert {\vec{\kappa}}\right\vert {^{2}}{{\left\vert {\vec{p}}\right\vert }^{2}}-2{p_{0}}\left\vert {\vec{\kappa}}\right\vert \left\vert {\vec{p}}\right\vert +{p^{2}}} & 0 & 0 & {\left\vert {\vec{\kappa}}\right\vert \left( {-{p_{0}}\left\vert {\vec{\kappa}}\right\vert \left\vert {\vec{p}}\right\vert +{p^{2}}}\right) } \\
0 & {-{p^{2}}+2{p_{0}}\left\vert {\vec{\kappa}}\right\vert \left\vert {\vec{p}}\right\vert } & 0 & 0 \\
0 & 0 & {-{p^{2}}+2{p_{0}}\left\vert {\vec{\kappa}}\right\vert \left\vert {\vec{p}}\right\vert } & 0 \\
{\left\vert {\vec{\kappa}}\right\vert \left( {-{p_{0}}\left\vert {\vec{\kappa}}\right\vert \left\vert {\vec{p}}\right\vert +{p^{2}}}\right) } & 0 & 0 & {\left\vert {\vec{\kappa}}\right\vert {^{2}p_{0}}^{2}+2{p_{0}}\left\vert {\vec{\kappa}}\right\vert \left\vert {\vec{p}}\right\vert -{p^{2}}}\end{array}\right). \label{oo1}
\end{equation}
\end{widetext}. The determinant provides one dispersion relation }%
\begin{equation}
\otimes (p)=p^{2}-2p_{0}|\vec{\kappa}||\vec{p}|=0,
\end{equation}%
{with multiplicity 4. The dimension of the null space of the matrix (%
\ref{oo1}) when this dispersion relation is satisfied is 3, }%
\begin{equation}
\dim \mathrm{Null}\left( O\big\vert_{\otimes (p)=0}\right) =3.
\end{equation}%
{It implies that we only can obtain three linearly independent (l.i.)
vectors which in principle would be the polarization vectors of the gauge
field. As one needs four polarization vectors to realize the Gupta-Bleuler
quantization, we conclude that the gauge condition, }$G(A_{\mu })=\partial
_{\mu }A^{\mu }+\kappa _{\mu \nu }\partial ^{\mu }A^{\nu }${, does
not work well in this case. A quick check of the failure of the Lorentz
gauge, }$G(A_{\mu })=\partial _{\mu }A^{\mu }${, can be made in the
case when the vectors }$\vec{p}${\ and }$\vec{\kappa}${\ are
perpendicular.}

{On the other hand, we can show that the condition }$G(A_{\mu
})=(\partial _{0}+\kappa ^{0j}\partial _{j})(A_{0}+\kappa
^{0k}A_{k})+\partial _{i}A^{i}${\ is a suitable choice for the
parity-odd and CPT-even electrodynamics. The momentum representation of the
operator defining the equation of motion of the gauge field now reads
\begin{widetext}
\begin{equation}
O^{\mu\nu}(p)=\left(
\begin{array}{cccc}
{\left\vert {\vec{\kappa}}\right\vert {^{2}}{{\left\vert {\vec{p}}\right\vert }^{2}}-2{p_{0}}\left\vert {\vec{\kappa}}\right\vert \left\vert {\vec{p}}\right\vert +{p^{2}}} & 0 & 0 & {\left\vert {\vec{\kappa}}\right\vert \left( {{{\left\vert {\vec{\kappa}}\right\vert }^{2}}{{\left\vert {\vec{p}}\right\vert }^{2}}-2{p_{0}}\left\vert {\vec{\kappa}}\right\vert \left\vert {\vec{p}}\right\vert +{p^{2}}}\right) } \\
0 & {-{p^{2}}+2{p_{0}}\left\vert {\vec{\kappa}}\right\vert \left\vert {\vec{p}}\right\vert } & 0 & 0 \\
0 & 0 & {-{p^{2}}+2{p_{0}}\left\vert {\vec{\kappa}}\right\vert \left\vert {\vec{p}}\right\vert } & 0 \\
0 & 0 & 0 & {2{p_{0}}\left\vert {\vec{\kappa}}\right\vert \left\vert {\vec{p}}\right\vert -{{\left\vert {\vec{\kappa}}\right\vert }^{2}}{{\left\vert {\vec{p}}\right\vert }^{2}}-{p^{2}}}
\end{array}\right). \label{oo2}
\end{equation}
\end{widetext}where we have also considered the vectors }$p_{i}${\
and }$\kappa _{i}${\ to be parallel\ and along the positive }$z$%
{-axis. The determinant provides two dispersion relations }%
\begin{eqnarray}
\otimes (p) &=&p^{2}-2p_{0}|\vec{\kappa}||\vec{p}|=0, \\[0.3cm]
\boxdot (p) &=&p^{2}-2p_{0}|\vec{\kappa}||\vec{p}|+|\vec{\kappa}|^{2}|\vec{p}%
|^{2}=0,
\end{eqnarray}%
{both having multiplicity 2. The dimension of the null spaces of the
operator (\ref{oo2}),\ for each dispersion relation, is }%
\begin{eqnarray}
\dim \mathrm{Null}\left( O^{\mu \nu }\big\vert_{\otimes =0}\right) &=&2, \\%
[0.15cm]
\dim \mathrm{Null}\left( O^{\mu \nu }\big\vert_{\boxdot =0}\right) &=&2,
\end{eqnarray}%
{respectively. So, every null space provides two l.i. vectors
totalizing four polarization vectors. Consequently, the gauge fixing
condition (\ref{gcc}) is a proper one to allow the implementation of the GB
quantization in\ this electrodynamics.}

\section{Remarks and conclusion}

We have studied the quantization of a CPT-even and parity-odd {LV }%
massless electrodynamics, performing the Gupta-Bleuler quantization, such as
it happens in the Maxwell electrodynamics under the Feynman gauge. We have
shown that the method can be implemented by using a gauge condition
depending explicitly\ on the LV coefficients, $G(A_{\mu })=(\partial
_{0}-\kappa _{j}\partial _{j})(A_{0}-\kappa _{k}A_{k})-\partial _{i}A_{i}$,
which appears as a LV modified version of the Lorenz gauge. Such a choice is
different from the one imposed to implement the GB quantization\ in the
CPT-even and parity-even LV electrodynamics, $G(A_{\mu })=\partial _{\mu
}A^{\mu }+\kappa _{\mu \nu }\partial ^{\mu }A^{\nu }$ ({see Ref. }%
\cite{GUPTAEVEN}). We also found the set of four\textit{\ l.i. }polarization
vectors that provided the sucessful implemention of the quantization.
{\ In Sec. \ref{app}, we explicitly argue that} if we use the
parity-even gauge condition for the parity-odd case, we can not establish a
set of linearly independent polarization vectors for arbitrary $\kappa _{i}$
and $p_{i}$. In other words, the parity-even gauge condition fails in
defining a set of polarization vectors when $\kappa _{i}$ and $p_{i}$ are
collinear. This problem is solved by the parity-odd gauge condition here
imposed.

\begin{acknowledgments}
We thank CAPES, CNPq and FAPEMA (Brazilian agencies) for partial financial
support.
\end{acknowledgments}


\begin{thebibliography}{99}
\bibitem{GUPTAEVEN} R. Casana, M. M. Ferreira Jr., and F. E. P. dos Santos,
Phys. Rev. D \textbf{90}, 105025 (2014).

\bibitem{Colladay} D. Colladay, V. A. Kostelecky, Phys. Rev. D \textbf{55},
6760 (1997); D. Colladay, V. A. Kostelecky Phys. Rev. D \textbf{58}, 116002
(1998); S. R. Coleman and S. L. Glashow, Phys. Rev. D \textbf{59}, 116008
(1999).

\bibitem{Jackiw} S.M. Carroll, G.B. Field and R. Jackiw, Phys. Rev. D
\textbf{41}, 1231 (1990).

\bibitem{Mewes} V. A. Kostelecky and M. Mewes, Phys. Rev. Lett. 87, 251304
(2001); V. A. Kostelecky and M. Mewes, Phys. Rev. D\ 66, 056005 (2002); V.
A. Kostelecky and M. Mewes, Phys. Rev.\ D 80, 015020\ (2009); V. A.
Kostelecky and M. Mewes, Phys. Rev. Lett. 97, 140401 (2006).

\bibitem{Soldati} A.A. Andrianov and R. Soldati, Phys. Rev. D 51, 5961
(1995); Phys. Lett. B 435, 449 (1998); A.A. Andrianov, R. Soldati and L.
Sorbo, Phys. Rev. D 59, 025002 (1998); A. A. Andrianov, D. Espriu, P.
Giacconi, R. Soldati, J. High Energy Phys. 0909, 057 (2009); J. Alfaro, A.A.
Andrianov, M. Cambiaso, P. Giacconi, R. Soldati, Int. J. Mod.Phys. A 25,
3271 (2010); V. Ch. Zhukovsky, A. E. Lobanov, E. M. Murchikova, Phys. Rev.
D73 065016, (2006); W. F. Chen and G. Kunstatter, Phys. Rev. D 62, 105029
(2000).

\bibitem{Adam} C. Adam and F. R. Klinkhamer, Nucl. Phys.\ B 607, 247 (2001);
C. Adam and F. R. Klinkhamer, Nucl. Phys.\ B \textbf{657}, 214 (2003).

\bibitem{Baeta} A. P. Baeta Scarpelli, H. Belich, J. L. Boldo, J. A.
Helayel-Neto, Phys. Rev. D \textbf{67}, 085021 (2003).

\bibitem{propag} R. Casana, M. M. Ferreira Jr, A. R. Gomes, F. E. P. dos
Santos, Phys. Rev. D \textbf{82}, 125006 (2010).

\bibitem{Massivephotons} M. Cambiaso, R. Lehnert, R. Potting, Phys. Rev. D
\textbf{85}, 085023 (2012).

\bibitem{Schreck} M. Schreck, Phys. Rev. D \textbf{86}, 065038 (2012).

\bibitem{Schreck2} M. Schreck, Phys. Rev. D \textbf{89}, 085013 (2014).

\bibitem{Hohensee} M. A. Hohensee, D. F. Phillips, R. L. Walsworth, \emph{%
Covariant quantization of Lorentz-violating electromagnetism},
arXiv:1210.2683.

\bibitem{colladay2014gupta} D. Colladay, P. McDonald, and R. Potting, Phys.
Rev. D \textbf{89}, 085014 (2014)

\bibitem{Colladaycov} D. Colladay, \emph{Covariant photon quantization in
the SME}, arXiv:1309.5890.

\bibitem{Colladaym} D. Colladay and P. McDonald, Phys. Rev. D \textbf{93},
125007 (2016).

\bibitem{Colladaym2} D. Colladay, P. McDonald, J. P. Noordmans, and R.
Potting , \emph{Covariant Quantization of CPT-violating Photons},
arXiv:1610.00169.

\bibitem{data} V. A. Kostelecky, N. Russell, Rev. Mod. Phys. \textbf{83}, 11
(2011).

\bibitem{kappamunu} B. Altschul, Phys. Rev. Lett. 98, 041603 (2007).

\bibitem{Gupta} S. N. Gupta, Proc. Phys. Soc. A \textbf{63}, 681 (1950).

\bibitem{Bleuler} K. Bleuler, Helv. Phys. Acta \textbf{23}, 567 (1950).

\bibitem{NAKANISHI} N. Nakanishi, Prog. Theo. Phys. \textbf{38}, 881 (1967).
\end{thebibliography}
\end{document}